\documentclass[preprints,article,accept,pdftex,moreauthors]{Definitions/mdpi} 

\firstpage{1} 
\pubvolume{1}
\issuenum{1}
\articlenumber{0}
\pubyear{2023}
\copyrightyear{2023}
\datereceived{} 
\dateaccepted{} 
\datepublished{} 
\hreflink{https://doi.org/} 

\newcommand{\eref}[1]{Eq.~(\ref{#1})}
\newcommand{\fref}[1]{Figure~\ref{#1}}

\newcommand{\nnnl}{\nonumber\\}	

\Title{How to determine the branch points of correlation
functions in Euclidean space II: Three-point functions}

\TitleCitation{How to determine the branch points of correlation
functions in Euclidean space II: Three-point functions}

\Author{Markus Q. Huber $^{1,}$*, Wolfgang J. Kern $^{2,3,}$* and Reinhard Alkofer $^{2,}$*}

\AuthorNames{Markus Q. Huber, Wolfgang J. Kern and Reinhard Alkofer}

\AuthorCitation{Huber, M. Q.; Kern, W.J.; Alkofer, R.}

\address{%
$^{1}$ \quad Institut f\"ur Theoretische Physik, Justus-Liebig-Universit\"at Giessen, 35392 Giessen, Germany\\
$^{2}$ \quad Institute of Physics, University of Graz, NAWI Graz, Universit\"atsplatz 5, 8010 Graz, Austria\\
$^{3}$ \quad Institute of Mathematics and Scientific Computing, University of Graz, NAWI Graz, Heinrichstraße 36, 8010 Graz, Austria}

\corres{Correspondence: markus.huber@theo.physik.uni-giessen.de, w.kern@uni-graz.at, reinhard.alkofer@uni-graz.at}

\abstract{
The analytic structure of elementary correlation functions of a quantum field is relevant for the calculation of masses of bound states and their time-like properties in general.
In quantum chromodynamics, the calculation of correlation functions for purely space-like momenta has reached a high level of sophistication, but the calculation at time-like momenta requires refined methods.
One of them is the contour deformation method.
Here we describe how to employ it for three-point functions.
The basic mechanisms are discussed for a scalar theory, but they are the same for more complicated theories and are thus relevant, e.g., for the three-gluon or quark-gluon vertices of quantum chromodynamics.
Their inclusion in existing truncation schemes is a crucial step for investigating the analytic structure of elementary correlation functions of quantum chromodynamics and the calculation of its spectrum from them.
}

\begin{document}

\section{Introduction}

Quantum chromodynamics (QCD) has a rich spectrum, and there are still many open questions about it.
Functional methods are one of several nonperturbative methods that can be used to unravel its mysteries, see, e.g., \cite{Cloet:2013jya,Eichmann:2016yit,Eichmann:2020oqt,Huber:2021yfy} for results on baryons, mesons, tetraquarks and glueballs.
In recent years, much progress has been made in the calculation of elementary correlation functions using functional methods, see, e.g., \cite{Williams:2015cvx,Cyrol:2017ewj,Huber:2018ned,Huber:2020keu,Gao:2021wun,Pawlowski:2022oyq,Papavassiliou:2022wrb,Ferreira:2023fva} and references therein.
However, as far as top-down calculations, which start directly from the Lagrangian of QCD, are concerned, the most advanced calculational schemes for functional equations have been applied to space-like momenta only.
For time-like momenta, calculations are more challenging due to the necessary adaptation of the numerical methods.
Complementary lattice methods provide direct access to correlation functions only at space-like momenta, see \cite{Cucchieri:2007md,Bogolubsky:2009dc,Maas:2011se,Oliveira:2012eh,Aguilar:2019uob,Pinto-Gomez:2022brg} for some exemplary results.

For perturbative integrals, one can use the Landau conditions \cite{Landau:1959fi} to determine the branch points of a diagram.
They are typically derived using the Feynman parametrization for the propagators.
For dressed propagators, however, this is not a viable approach, and an analysis more along the lines of numerical calculations is required.

Such an approach to access the analytic properties of correlation functions is provided by the contour deformation method (CDM).
It deals with the intricacies introduced by time-like momenta by modifying the integration path in the integral appropriately.
This enables numeric calculations but also leads to insights into the analytic structure of correlation functions.
Originally it was devised for QED \cite{Maris:1995ns} for a special case and then subsequently generalized \cite{Alkofer:2003jj,Eichmann:2007nn,Windisch:2012zd,Windisch:2012sz,Strauss:2012dg,Windisch:2013dxa,Eichmann:2019dts,Fischer:2020xnb,Huber:2022nzs}.
Other direct methods include the shell method \cite{Fischer:2008sp}, the use of the Cauchy-Riemann equations \cite{GimenoSegovia:2008sx}, the covariant spectator theory framework \cite{Biernat:2018khd}, the Cauchy method \cite{Fischer:2005en,Krassnigg:2009gd}, or spectral representations including the Nakanishi integral representation \cite{Nakanish:i1971gtf,Sauli:2001mb,Sauli:2006ba,Jia:2017niz,Frederico:2019noo,Horak:2020eng,Mezrag:2020iuo,Horak:2021pfr,Horak:2022myj,Duarte:2022yur,Horak:2022aza}.

Here we summarize the case of three-point functions systematically.
In particular, we aim at an accessible description of the method in the spirit of \cite{Windisch:2013dxa} of which this article can be considered a follow-up, hence the title.
Calculational details, while mathematically straightforward, can be found elsewhere \cite{Huber:2022nzs}.
First, we introduce the basic idea with the example of a two-point integral in Sec.~\ref{sec:twoPoint}.
As a new feature, we pay particular attention to the possibility of deforming not only the integration contour in the radial variable but also in an angle, thereby deforming the branch cuts that require a deformation of the radial variable in the first place in Sec.~\ref{sec:angleDeform}.
For the three-point function in Sec.~\ref{sec:threePoint}, we start with simplified kinematics before we discuss the general case.

\section{Contour deformation method}

In the following we will work with propagators with a generic mass $m$.
The analysis is valid both perturbatively, where $m$ is the bare mass, but also nonperturbatively if the propagator features a single pole and $m$ is the corresponding mass.
Cuts can also be considered \cite{Huber:2022nzs}.

\subsection{Basic example: The two-point integral}
\label{sec:twoPoint}

\begin{figure*}[b]
\includegraphics[width=0.3\textwidth]{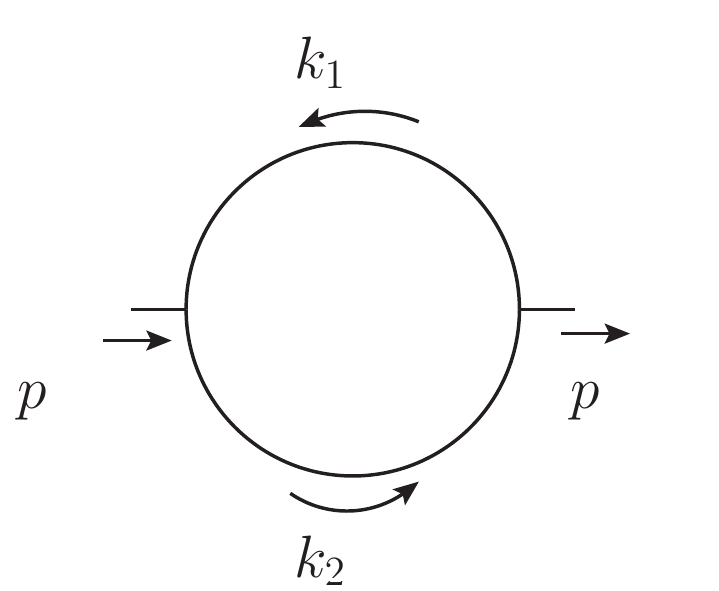}\hfill
\includegraphics[width=0.3\textwidth]{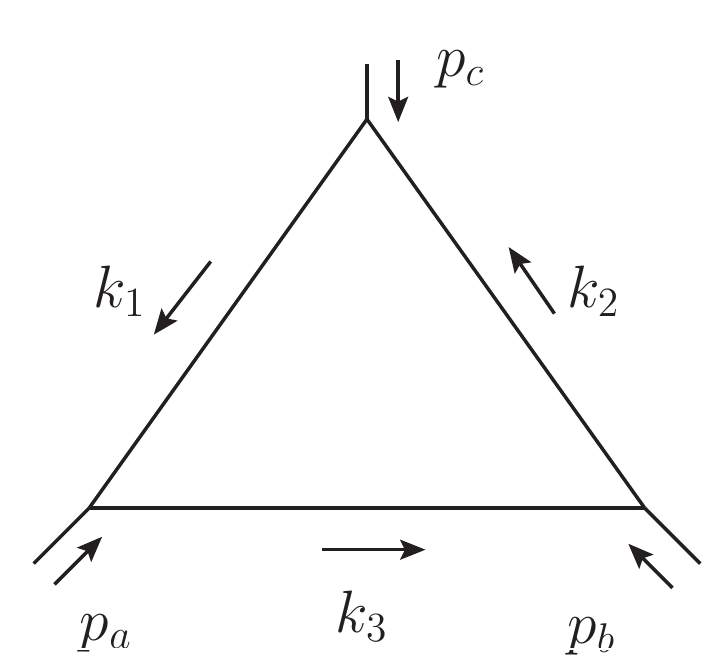}\hfill
\includegraphics[width=0.3\textwidth]{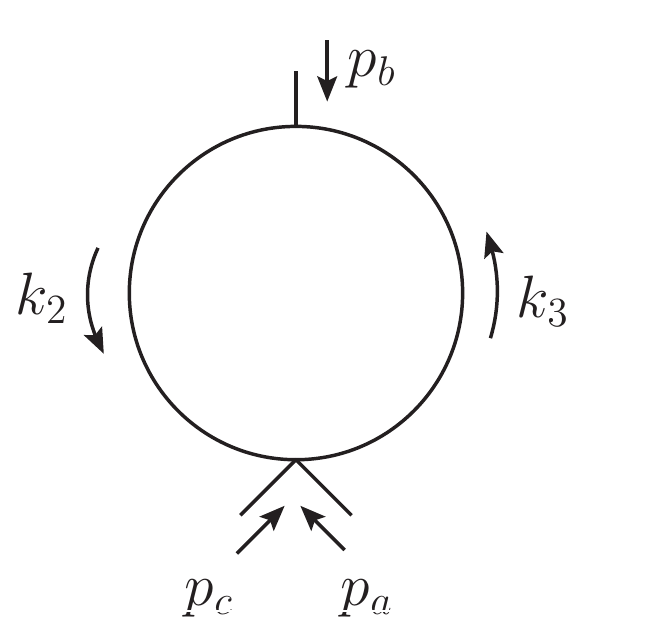}
\caption{Momentum routing for the propagator's one-loop selfenergy (left), the triangle diagram (center) and the swordfish diagram (right).
The internal momenta $k_i$ are combinations of external and loop momenta, see Eqs.~(\ref{eq:int_I2}) and (\ref{eq:int_f}).}
\label{fig:LC_routing}
\end{figure*}

For illustration purposes we consider the Euclidean one-loop two-point integral
\begin{align}\label{eq:int_I2}
 I_2(p^2)=\int \frac{d^dq}{(2\pi)^d} \frac{1}{q^2+m^2}\frac{1}{(q-p)^2+m^2},
\end{align}
see \fref{fig:LC_routing}.
For conciseness, we fix the number of dimensions to four in the following, but the generalization is straightforward.
Using hyperspherical coordinates, two angles can be integrated out, and the radial variable $r=\sqrt{q^2}$ as well as one angle $\theta_1$ remain.
The integrand has poles at $q^2=-m^2$ and $(q-p)^2=r^2+p^2-2\sqrt{p^2}\,r\cos\theta_1-m^2$ which must not be crossed during the integration.
Performing the angle integration first, the second propagator leads to branch cuts corresponding to the integration of $\theta_1$ from $0$ to $\pi$.
It can be parameterized as ($z_1=\cos \theta_1$)
\begin{align}
\label{eq:denom_sol}
\gamma_{\pm}(z_1;p^2,m^2)=&\sqrt{p^2}\,z_1 \pm i\sqrt{m^2+p^2(1-z_1^2)} \nnnl
 =& \sqrt{p^2}\cos\theta_1 \pm i\sqrt{m^2+p^2\sin^2\theta_1},
\end{align}
which is obtained by solving the quadratic equations $(q-p)^2=-m^2$ for $r$.
The analytic structure of the remaining integrand in $r$ thus consists of the poles at $\pm i\,m$ and these two cuts.
The dependence on the external momentum $p$ enters via the latter.
We stress that we use the radial variable $r=\sqrt{q^2}$ instead of $q^2$ to avoid ambiguities later for the three-point function \cite{Huber:2022nzs}.

The integration of $r$ starts at $0$ and ends at the chosen UV cutoff.
When the external momentum $p^2$ is positive, neither the poles nor the cuts interfere, and the integration can be performed along the real axis.
This changes when $p^2$ is complex or negative.
An example is shown in \fref{fig:sing_prop}.
For the chosen value of $\sqrt{p^2}$, the cut crosses the real axis.
To avoid crossing the cut, the contour of the $r$ integration needs to be deformed.
A simple choice is to integrate along a straight line from the origin to $\sqrt{p^2}$ and further out until a chosen stopping point.
From there, the integration can be closed by continuing in an arc to the UV cutoff, see \cite{Fischer:2020xnb} for details on a systematic implementation called ray method.

Analyticity of the integral means that it can be calculated in a neighborhood of a point by continuously deforming the involved integration contours.
If this is not the case, a nonanalyticity is found.
This happens in this example exactly then when an endpoint of a cut from one propagator touches the pole of the other propagator.
The deformation on the $r$ integration contour would then need to jump over the pole, thereby picking up its residue.
The condition that the endpoint of the cut touches the pole is \cite{Windisch:2013dxa}
\begin{align}\label{eq:cond_2p}
 \gamma_{\pm}(\pm1;p^2,m^2)=\pm i\,m.
\end{align}
The only nontrivial solutions leads to $p^2=-4m^2$ which is the result expected from the Landau conditions \cite{Landau:1959fi}.

\subsection{Deformations of the angle integration contour}
\label{sec:angleDeform}

\begin{figure}[tb]
\centering
\includegraphics[width=0.48\textwidth]{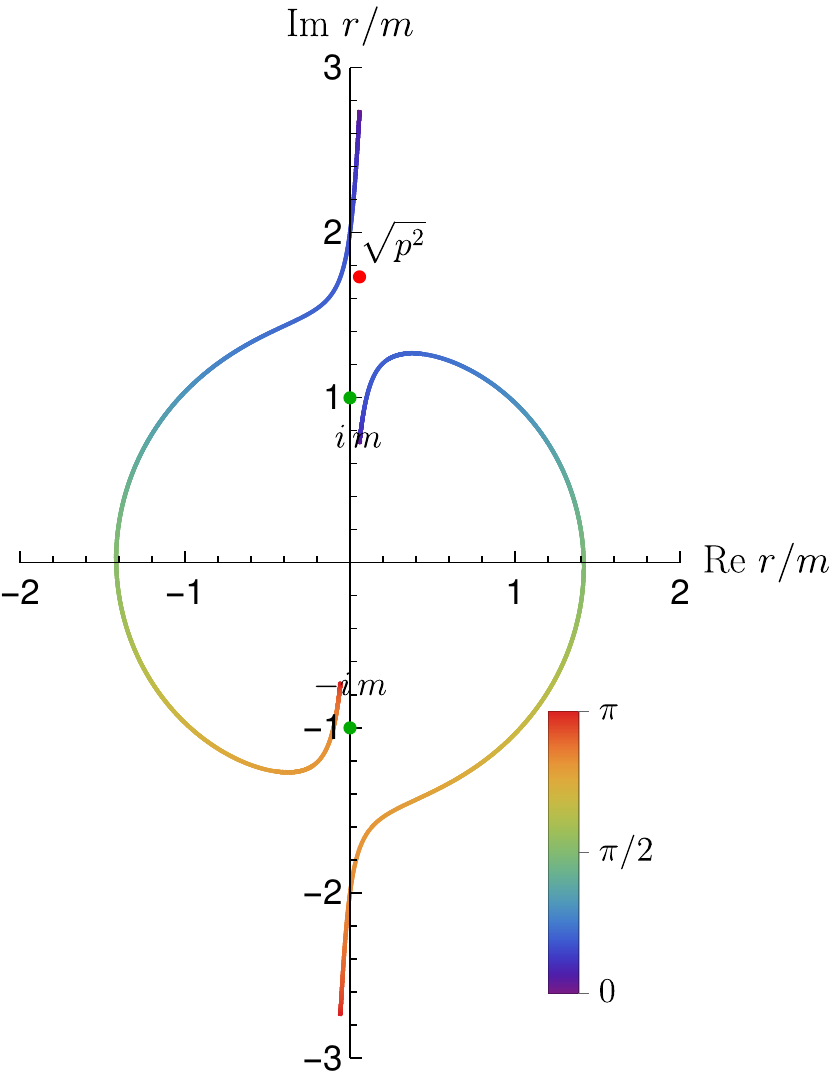}\hfill
\includegraphics[width=0.48\textwidth]{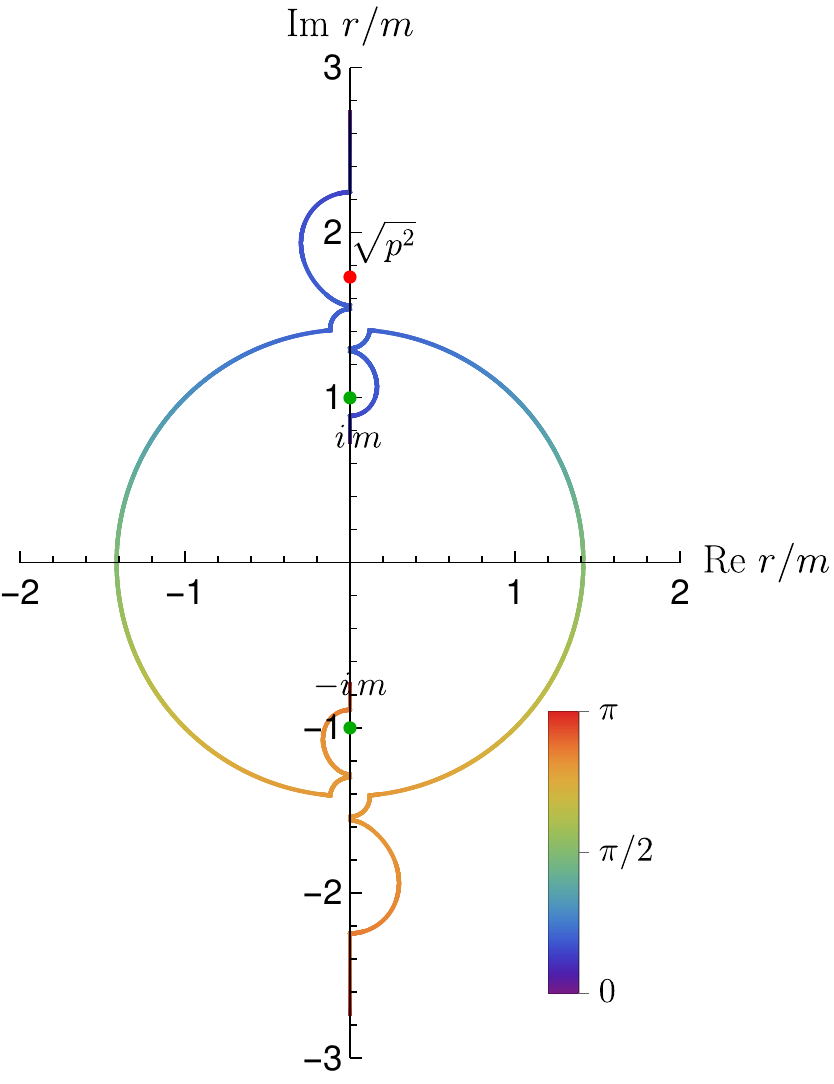}
\caption{Examples for the singularity structure $\gamma_{\pm}(z_1;p^2,m^2)$ of the propagator $r=\sqrt{q^2}$ integrand for $p^2=(-3+0.2i)m^2$ (left) and $p^2=-3m^2$ (right).
The lines denote branch cuts stemming from the angular integral with the value of the angle $\theta_1$ indicated by the color.
In the left plot, the angle integral is performed in a straight line from $0$ to $\pi$.
In the right plot, the angle integral is modified such as to avoid the poles at $\pm i\,m$ and gaps are opened at $\pm\sqrt{-2}m$ using the path of \eref{eq:t1_complex}.
The two bulges at $\pm 2\,i\,m$ are a consequence of deforming $\theta_1$ around $\pm i\,m$.
The green dots are the poles from the second propagator.
The red dots indicate the external $p^2$ and are only plotted for reference.
}
\label{fig:sing_prop}
\end{figure}

Based on the left plot in \fref{fig:sing_prop} one might wonder what happens when $p$ moves onto the imaginary axis, or, equivalently, when $p^2$ is negative.
Does the branch cut then go over the pole?
If the angular integration is performed in a straight line from $0$ to $\pi$, it indeed does, but we can also deform the integration contour  for the angular integral.
An example of this is shown in the right plot of \fref{fig:sing_prop}.
Two observations should be made.
First, this explains the relevance of the endpoints for \eref{eq:cond_2p} because they cannot be moved around in contradistinction to the integration path between endpoints.
Second, if the contour is deformed to avoid a pole in one place, it introduces deformations elsewhere as well.

When $p^2$ is real and negative, the following values of $\theta_1$ need extra care.
First, it is possible that the pole lies on the branch cut.
This happens at
\begin{align}
 \theta^{*p\pm}_{1}=\arccos\left(\pm i\frac{\sqrt{p^2}}{2m}\right).
\end{align}
Second, the two cuts touch each other on the imaginary axis at 
\begin{align}
 \theta^{*c\pm}_1=\arcsin\left(\pm i\frac{m}{\sqrt{p^2}}\right).
\end{align}

\begin{figure}[tb]
 \begin{center}\includegraphics[width=0.49\textwidth]{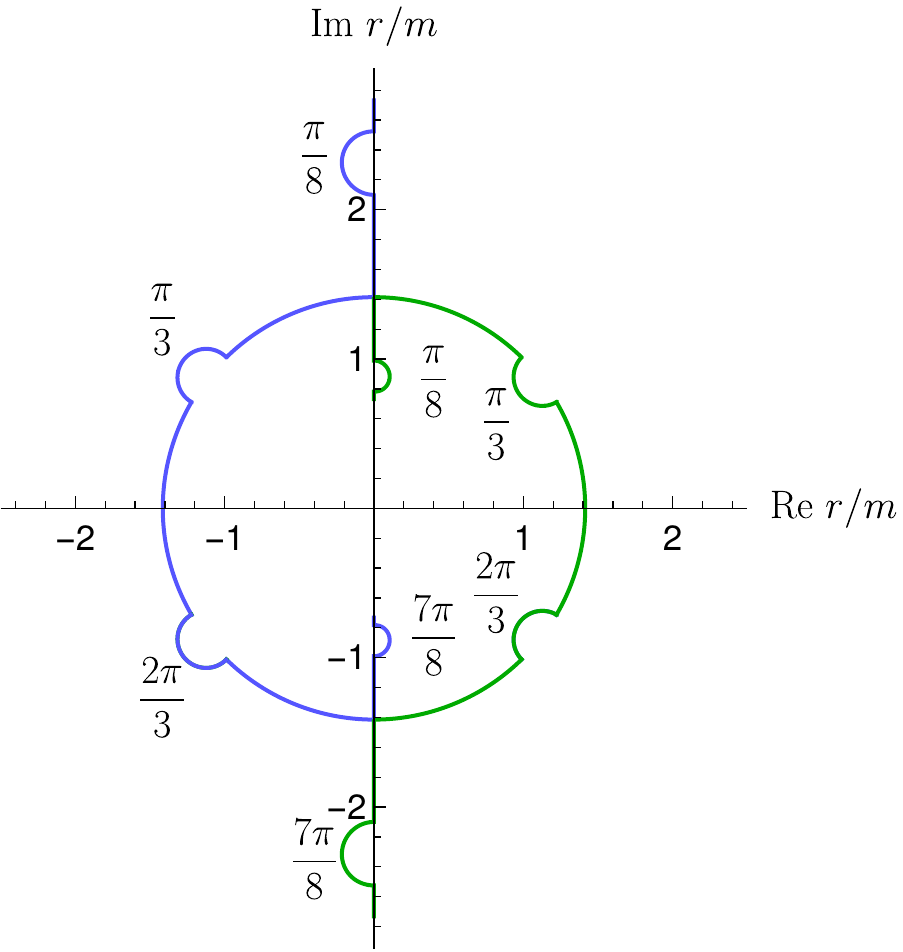}\end{center}
 \caption{Bulges from deforming the angle integration in $\theta_1$ via $\theta_1^+$ as given in \eref{eq:t1_complex} for the indicated values of $\theta^*_1$.}
 \label{fig:bulges}
\end{figure}

A concrete example of how to avoid a point $\theta_1^*$ in the integration of $\theta_1$ from $0$ to $\pi$ is in the form of a semicircle with radius $s$:
\begin{align}
 \theta_1\rightarrow \theta_{1}^\pm=
 \begin{cases}
    \theta^*_{1}+s\,e^{\pm i\frac{\theta_1-\theta^*_1-s}{s}\frac{\pi}{2}} &\quad  |\theta_1-\theta^*_{1}|<s \\
    \theta_1 &\quad  \text{otherwise}
 \end{cases}
 \label{eq:t1_complex}
\end{align}
Note the free choice of the sign of the phase which corresponds to two directions the corresponding deformations in the $r$ plane can have.
For this reparametrization the pattern of evasive bulges in the complex $r$ plane depicted in \fref{fig:bulges} emerges.

\subsection{The triangle integral}
\label{sec:threePoint}

We turn now to three-point functions.
They can have two different diagrams, a swordfish diagram and a triangle diagram, see \fref{fig:LC_routing}.
By choosing the momentum routing appropriately, the former can be described on the same footing as the two-point integral.
Hence, we only discuss the triangle diagram in the following.

The triangle diagram with external momenta $p_a$, $p_b$ and $p_c=-p_a-p_b$ has the following form:
\begin{align}\label{eq:int_q2_f}
 I_3(p_a,p_b,p_c)=\int dr\, r^{3}f(q,p_a,p_b,p_c)
\end{align}
with
\begin{align}\label{eq:int_f}
 f(q,p_a,p_b,p_c)=&\int (\sin\theta_1)^{2} d\theta_1\int \sin\theta_2 d\theta_2 \frac{1}{q^2+m^2}\frac{1}{(q-p_a)^2+m^2}\frac{1}{(q+p_b)^2+m^2}.
\end{align}
With the chosen routing, one propagator creates poles at $p=\pm i\,m$ and the other two cuts of the form
\begin{subequations}
\label{eq:gamma_ab}
\begin{align}
\gamma_{a\pm}(z_1;p_a^2,m^2)&=\gamma_{\pm}(z_1;p_a^2,m^2),\\
\gamma_{b\pm}(\tilde z;p_b^2,m^2)&=\gamma_{\pm}(-\tilde z;p_b^2,m^2)
\end{align}
\end{subequations}
with
\begin{align}\label{eq:thetatilde}
\tilde z=\cos \tilde \theta=\cos{\theta}\,\cos{\theta_1}+\sin{\theta}\,\sin{\theta_1}\, \cos{\theta_2}.
\end{align}
The analysis of the analytic structure of the $r$ plane is more complicated than that of the two-point integral due to the existence of twice as many cuts and the appearance of a second angle integral.
As it is instructive and already shows the basic features, we will in the following discuss a simplified case first.

\subsubsection{Restricted kinematics}

We restrict the kinematics by setting $p_b^2=p_a^2=p^2$ and consequently $p_c^2=2p^2(1+\cos\theta)$.
As a consequence, the cuts $\gamma_{b\pm}$ are a subset of $\gamma_{a\pm}$.
From the two-point integral we know that a branch point in the external momentum arises when a cut in $r=\sqrt{q^2}$ cannot be deformed such as to avoid the pole.
This happened for the end points of the cuts.
Here, new possibilities arise because of the three present propagators.
Deformations of integration contours now need to respect constraints from all three of them.
Only one propagator depends on the angle $\theta_2$.
In that case, the end points $\theta_2=0$ and $\theta_2=\pi$ are relevant as they are fixed whereas we could perform an additional deformation of the $\theta_2$ integration in between.
For conciseness, we work with $\theta_2=0$ in the following discussion.

\begin{figure*}[t]
\begin{minipage}{0.49\textwidth}
\includegraphics[width=0.98\textwidth]{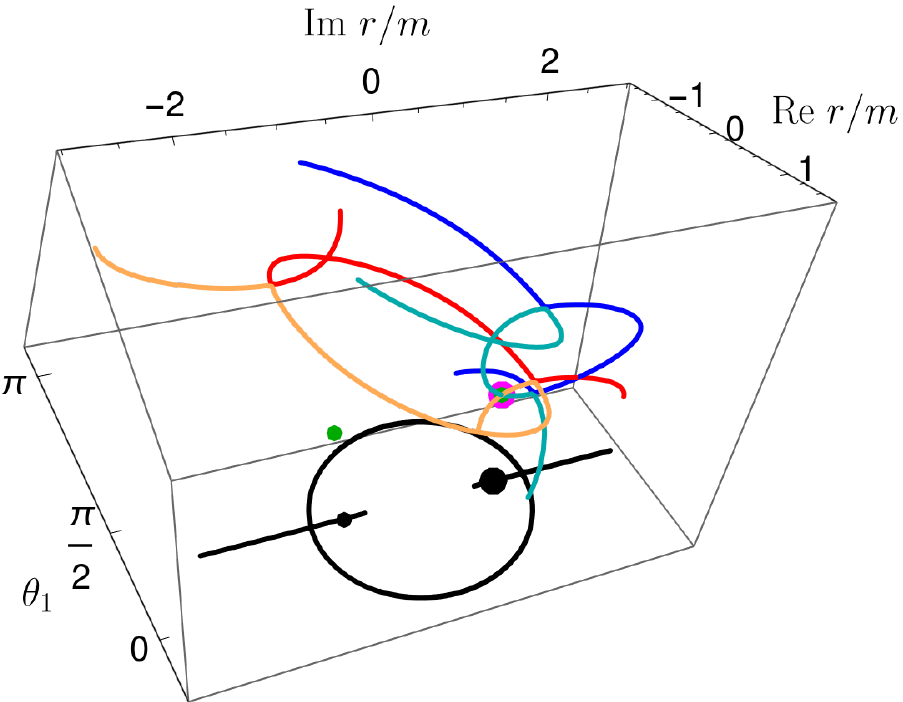}
\end{minipage}
\hfill
\begin{minipage}{0.41\textwidth}
\includegraphics[width=0.98\textwidth]{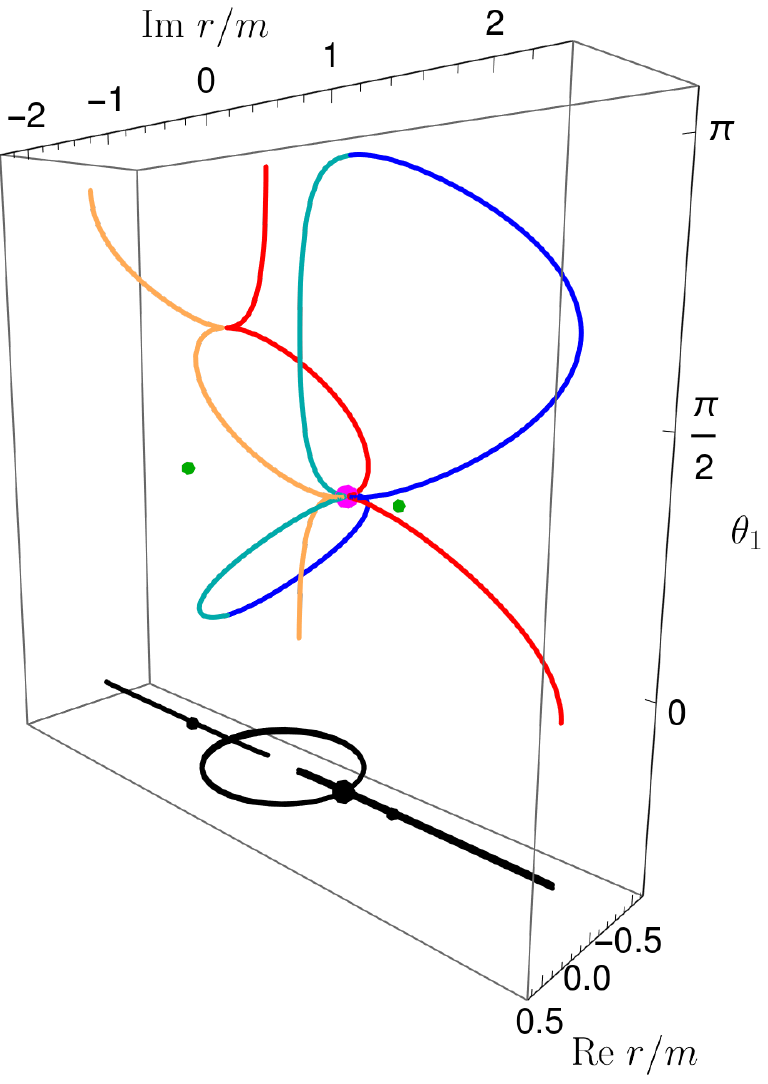}
\end{minipage}
\caption{
Examples for the singularity structure of the triangle $r=\sqrt{q^2}$ integrand.
The lines denote branch cuts stemming from the $\theta_1$ angle integral.
The red line is $\gamma_{a+}$, the orange line $\gamma_{a-}$, the blue line $\gamma_{b+}$, and the cyan line $\gamma_{b-}$.
The green dots are the poles from a propagator, the magenta ones indicate where the relevant crossings of cuts/poles are.
Left: $p^2=-3m^2$, $\theta=2\pi/3$, $\theta_2=\pi$, two cuts cross at $i\,m$ so the magenta dot is at the same point as a green one.
Right: $p^2=-4m^2/3$, $\theta=\pi/3$, $\theta_2=\pi$, four cuts touch at $-m^2/3$.
The black lines are projections of the cuts into one plane.
}
\label{fig:sing_vert_3d}
\end{figure*}

As already mentioned above, the cuts created by the propagators lie on top of each other.
However, the important observation is that for a given value of the angle $\theta_1$ they do not necessarily agree.
To illustrate this, we add a third axis for $\theta_1$ in the plots of the branch cuts.
Two examples are depicted in \fref{fig:sing_vert_3d}.
There, one can see the four branch cuts (two from each propagator) and the points where cuts from different propagators cross.
This happens when
\begin{align}\label{eq:sol_t1_1}
 \theta_{1,c\pm}=\frac{\theta}{2}+\frac{\pi}{2},
\end{align}
which is the solution of the condition that the two propagators agree: 
\begin{align}\label{eq:3p_twoCut_cond}
 -\cos \theta_1 = \cos \theta\cos\theta_1 + \sin\theta\sin\theta_1 \, .
\end{align}

The left plot in \fref{fig:sing_vert_3d} corresponds to the case when the two cuts meet at a pole of the third propagator.
Plugging \eref{eq:sol_t1_1} into the corresponding condition that the first cut agrees with the pole $i\,m$,
\begin{align}
 \gamma_{a,+}(z_1;p^2,m^2)=i\,m,
\end{align}
leads to the following branch point in $p^2$:
\begin{align}\label{eq:sol_p1}
 p_{B,1}^2&=-4m^2\cos^2\left(\frac{\theta}{2}+\frac{\pi}{2}\right) = -4m^2 \sin^2\frac{\theta}{2}.
\end{align}
One can convince oneself with the help of \fref{fig:bulges} that it is not possible to deform the angle integration in $\theta_1$ such as to open a gap around the pole because each branch cut requires a different sign of the phase in \eref{eq:t1_complex}.

\begin{figure*}[t]
 \includegraphics[width=0.49\textwidth]{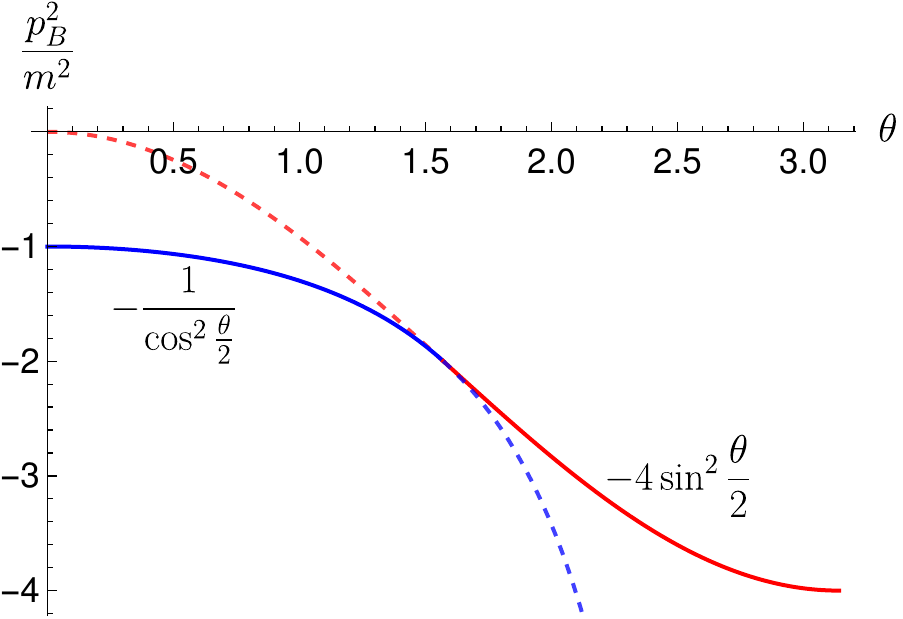}\hfill
 \includegraphics[width=0.49\textwidth]{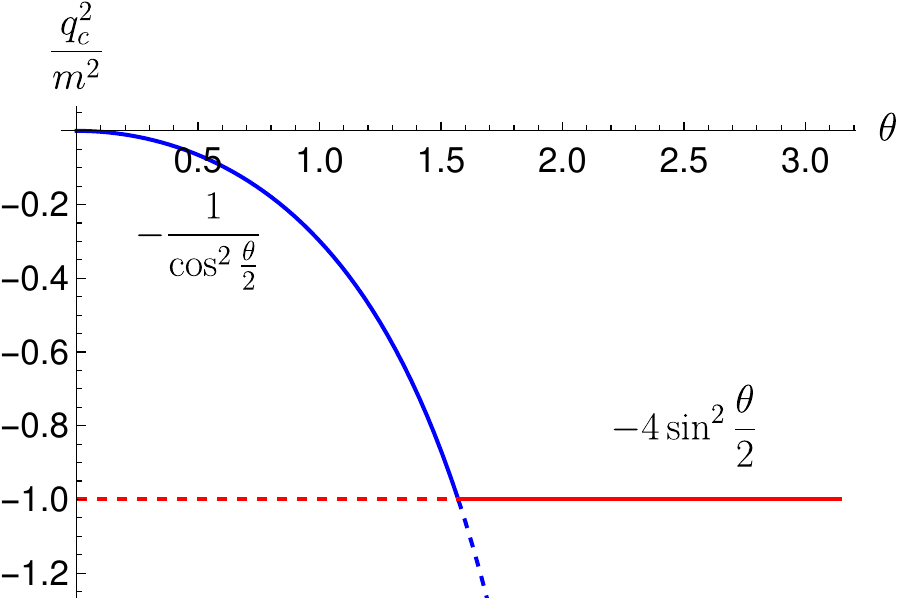}
 \caption{Left: The positions of the two potential branch points from Eqs.~(\ref{eq:sol_p1}) and (\ref{eq:sol_p2}).
 Right: The critical points as functions of $\theta$.
 The dashed lines correspond to the irrelevant cases and the continuous ones to the physical solutions.}
 \label{fig:twoSols}
\end{figure*}

When $p^2>-2m^2$, the pole lies outside of the semicircle parts of the branch cuts and does not interfere.
However, for certain values of the \textit{external} angle $\theta$, the four cuts meet on the imaginary axis, see the right plot in \fref{fig:sing_vert_3d}.
Again, there is no deformation possible and a branch point is created at this $p^2$.
The four cuts meet at the same value of $\theta_1$ when the second term in \eref{eq:denom_sol} vanishes and $\theta_1$ is given by \eref{eq:sol_t1_1}.
This leads to the branch point
\begin{align}\label{eq:sol_p2}
 p_{B,2}^2 &=-\frac{m^2}{\sin^2\left(\frac{\theta}{2}+\frac{\pi}{2}\right)}=-\frac{m^2}{\cos^2\frac{\theta}{2}}.
\end{align}
The corresponding point in the $r=\sqrt{q^2}$ plane is
\begin{align}\label{eq:qc2}
 q_{c,2}^2=\gamma_{a\pm}^2(\theta_{1,c};p_{B,2}^2,m^2)=-m^2 \tan^2\frac{\theta}{2}.
\end{align}

The two potential branch points $p_{B,1}^2$ and $p_{B,2}^2$ are shown in the left plot of \fref{fig:twoSols}.
Since $p_{B,1}^2>p_{B,2}^2$, one might think that $p_{B,1}^2$ is the relevant branch point, but decisive is which singular point appears closer to the origin in the $r$ plane.
By singular point we refer to the value of $r$ which forbids the contour deformation.
In the first case, this is $q_{c,1}^2=-m^2$, and in the second one $q_{c,2}^2$ from \eref{eq:qc2}.
They are plotted as a function of $\theta$ in \fref{fig:twoSols}.
For $\theta\leq\pi/2$, we have $q_{c,2}^2>q_{c,1}^2$.
When starting with $p^2$ at zero and then decreasing it (see left plot in \fref{fig:twoSols}), the branch points in the $r$ plane do not pose a problem as long as $-m^2< p^2$ because no circular parts crossing the real line are created.
When $p^2\leq-m^2$, the singular point $q_{c,2}^2$ forbids deforming the contour if $\theta\leq\pi/2$ and the singular point $q_{c,1}^2$ if $\theta\geq \pi/2$ (see right plot in \fref{fig:twoSols}).

\begin{figure*}[t]
 \includegraphics[width=0.48\textwidth]{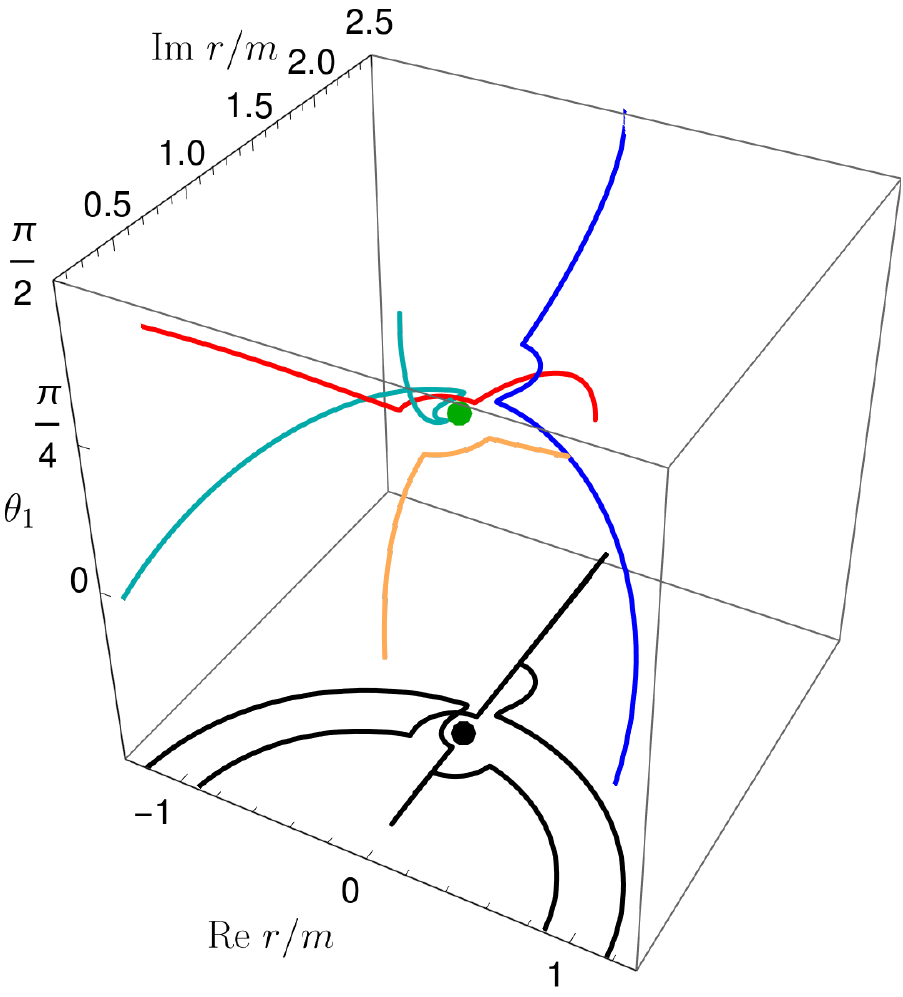}
 \hfill
 \includegraphics[width=0.48\textwidth]{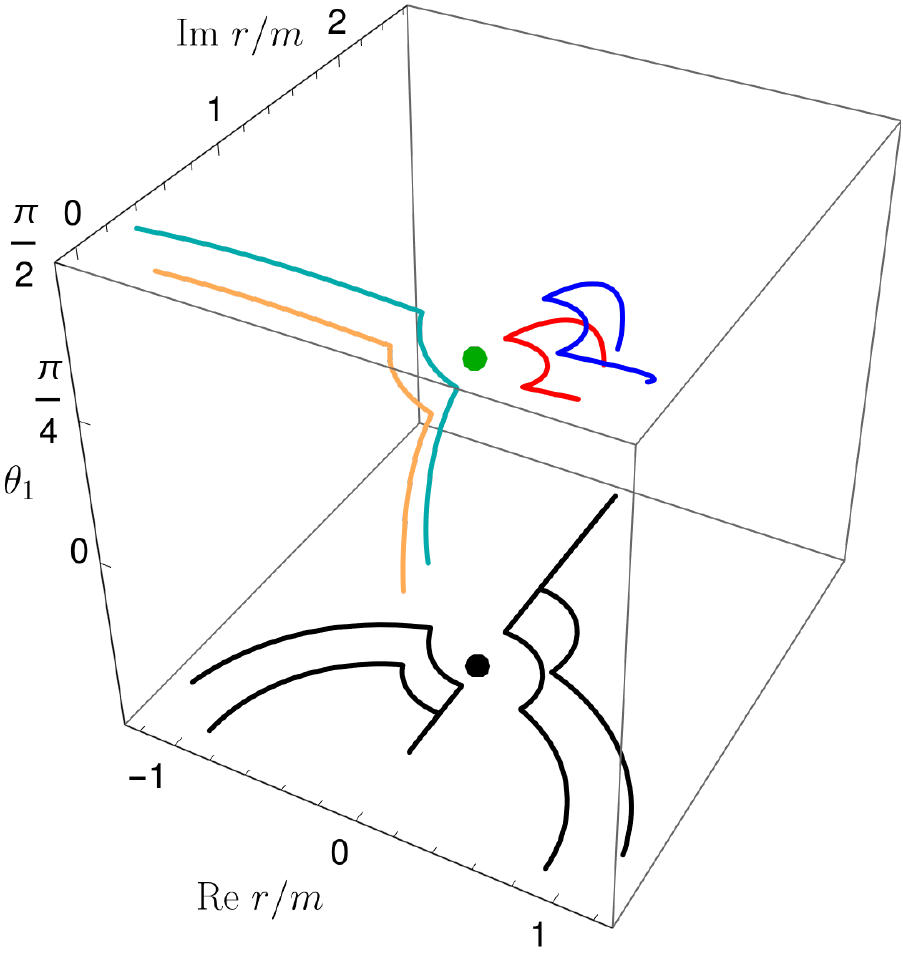}
 \caption{The cuts for $p_a^2=-1.8m^2$, $p_b^2=-2.4m^2$ and $\theta_1 \in [0,\pi/2]$.
 In the left/right plot, $\theta_+/\theta_-$ is used.
 The contours are deformed around the point where the cuts touch.
 This opens a path for $\theta_-$ but not for $\theta_+$ as can be seen in the projections (black).
 Colors as in \fref{fig:sing_vert_3d}.
 }
 \label{fig:cp_cm}
\end{figure*}

There is a simple way of finding the branch point $p_{B,2}^2$.
Since only two propagators are involved, we can change the routing such that these two propagators have the momentum arguments $q$ and $q-p_i$ with $i=a,b,c$.
The analysis is then equivalent to the two-point integral and we obtain the branch point $p_i^2=-4m^2$.
If we do this for the chosen kinematic situation, we obtain $p_c^2=-4m^2$, which, in turn, leads to $p^2=-2m^2/(1+\cos\theta)$.
This is equivalent to the solution found above.
While this is a direct and simple way to find the branch point, it is inconvenient for numeric calculations where one does not want to work with different momentum routings.

To summarize, we have the following solution for the branch point of the triangle diagram as a function of $\theta$ when $p_a^2=p_b^2=p^2$:
\begin{align}\label{eq:pB2_restKin}
p_B^2=\left \lbrace \begin{array}{c c} -4m^2 \sin\left(\frac{\theta}{2}\right)^2 & \quad \frac{\pi}{2}\leq \theta \leq \pi  \\ \frac{-m^2}{\cos\left(\frac{\theta}{2}\right)^2} & \quad  0\leq \theta \leq \frac{\pi}{2} \end{array} \right. .
\end{align}

\subsubsection{General kinematics}

We now remove the restriction on the kinematics and discuss the case for $p_a^2\neq p_b^2$.
Again we have to distinguish between the case when both branch cuts meet at the pole and the case without the pole.

For the first case, we know that $r$ must be equal to $\pm\,i\,m$.
We plug this into the other two propagators and equate their denominators.
This leads to the condition
\begin{align}\label{eq:genKinCond}
 p_a^2\, p_b^2\, p_c^2=&m^2(p_a^4+p_b^4+p_c^4-2(p_a^2\,p_b^2+p_a^2\,p_c^2+p_b^2\,p_c^2)),
\end{align}
where $p_c^2=p_a^2+p_b^2+2\sqrt{p_a^2}\sqrt{p_b^2}\cos \theta$ was used.
This equation has two solutions for $\theta$.
It remains to check if the contour deformations are possible or not.
As it turns out, they are only for one solution, see \fref{fig:cp_cm} for examples.
Thus, we have found one surface in the space spanned by $p_a^2$, $p_b^2$ and $p_c^2$ corresponding to a threshold:
\begin{align}
 p_{c+}^2=&\frac{1}{2m^2} \Big( 2(p_a^2+p_b^2)m^2 + p_a^2 \,p_b^2+ \sqrt{p_a^2}\sqrt{4m^2+p_a^2} \sqrt{p_b^2}\sqrt{4m^2+p_b^2} \Big).
\end{align}

\begin{figure*}
\includegraphics[width=0.48\textwidth]{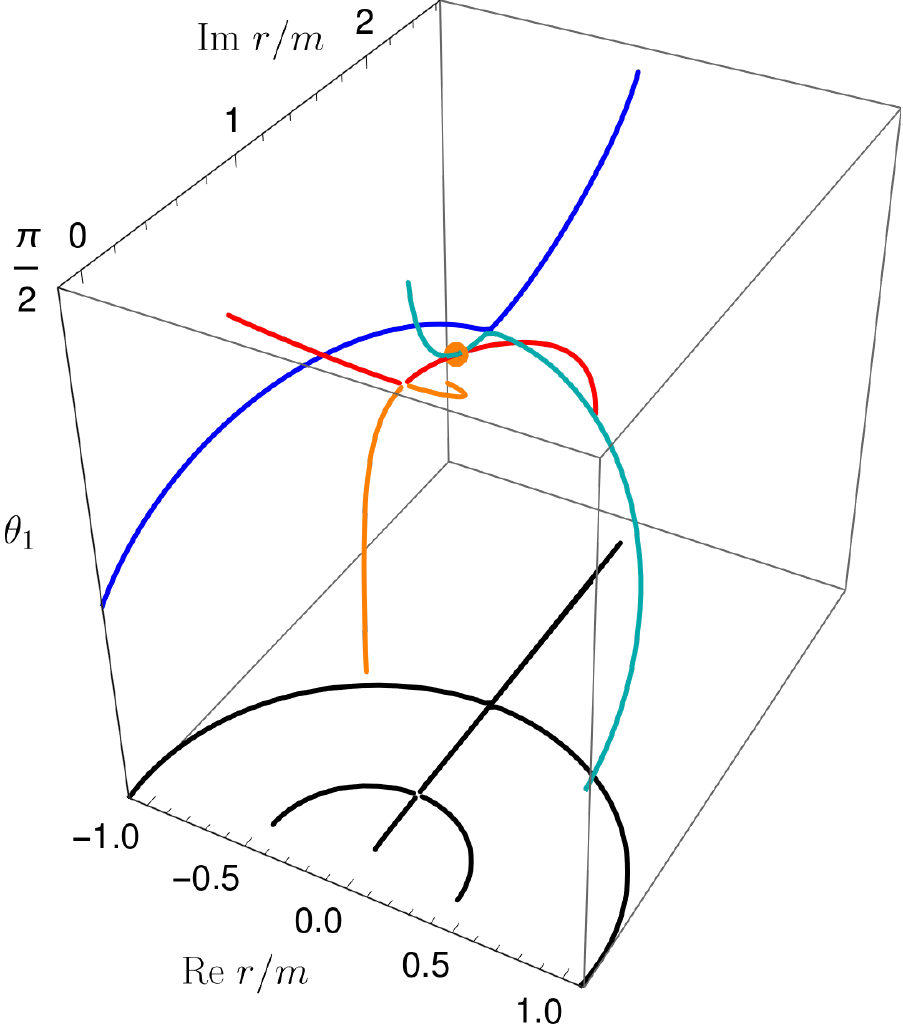}
\hfill
\includegraphics[width=0.48\textwidth]{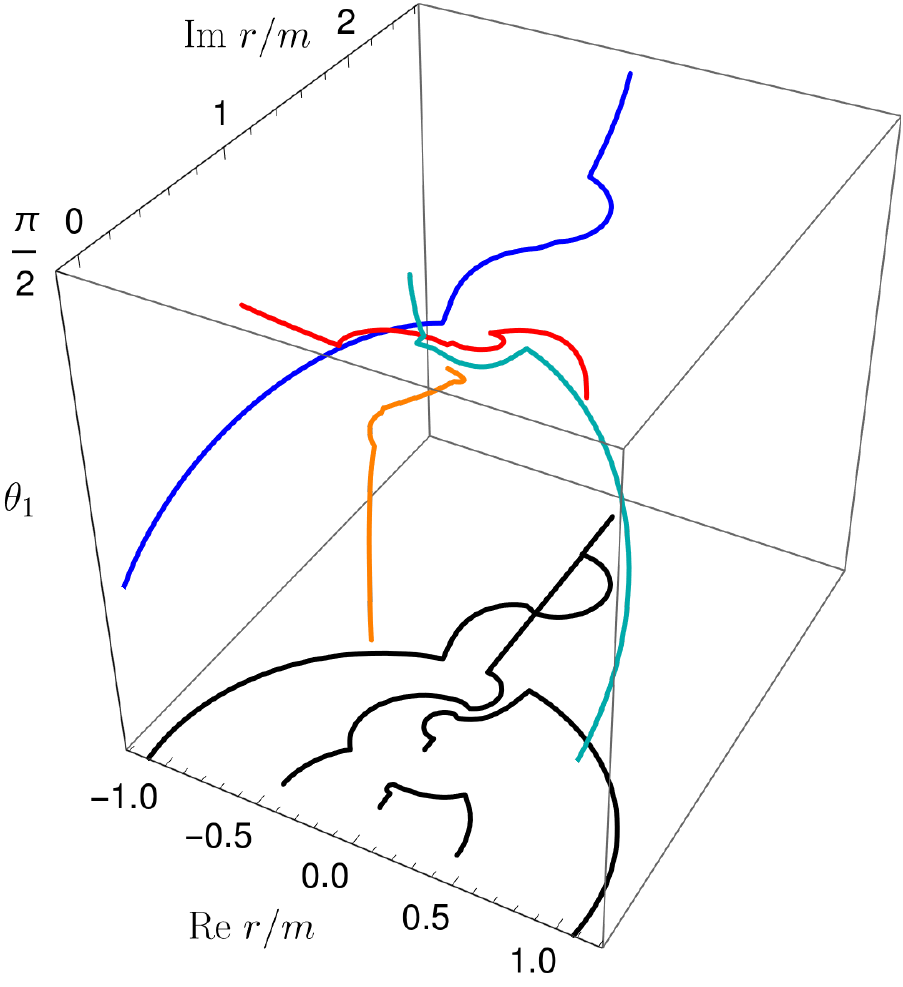}
\caption{Cuts of the triangle diagram for $p_a^2=-1.2m^2$, $p_b^2=-2.2m^2$ and $\theta_1 \in [0,\pi/2]$.
In the left plot, $p_c^2=-4m^2$, and the cuts touch at $\sqrt{(p_a^2+p_b^2)/2+m^2}$.
In the right plot, the value for $\theta$ is slightly shifted compared to the left plot and the cuts cross at two points.
However, a contour deformation can be found that allows to lead the integration out of the two circles as can be seen by the projected cuts in black.
Colors as in \fref{fig:sing_vert_3d}.
}
\label{fig:vert_abm2}
\end{figure*}

For the second situation one can directly obtain the thresholds by considering all possible pairs of propagators and adapting the routing to have the arguments $q$ and $q-p_i$, $i=a,b,c$.
This leads to the thresholds:
\begin{subequations}
\begin{align}
 p_a^2=&-4m^2,\\
 p_b^2=&-4m^2,\\
 p_c^2=&-4m^2,
\end{align}
\end{subequations}
which correspond to walls in the space spanned by $p_a^2$, $p_b^2$ and $p_c^2$.
To find this result without changing the routing, we need to find the singular point creating the branch point.
It turns out that this point is where the inner straight parts of the two cuts touch.
We can determine that by choosing two values for $p_a^2$ and $p_b^2$ and plugging in the value for $\theta$ determined from $p_c^2=-4m^2$.
If the cuts touch before the pole at $-m^2$, this creates a branch point.
If $\theta$ is changed, the two cuts either do not touch or cross at two points.
Exemplary situations are depicted in \fref{fig:vert_abm2} where it is also shown that a contour deformation can be found in the latter case.
In the case where they only touch, no deformation is possible because there is only one critical point which would require opposite directions of the deformations for each cut.

\begin{figure}
\begin{center}\includegraphics[width=.49\textwidth]{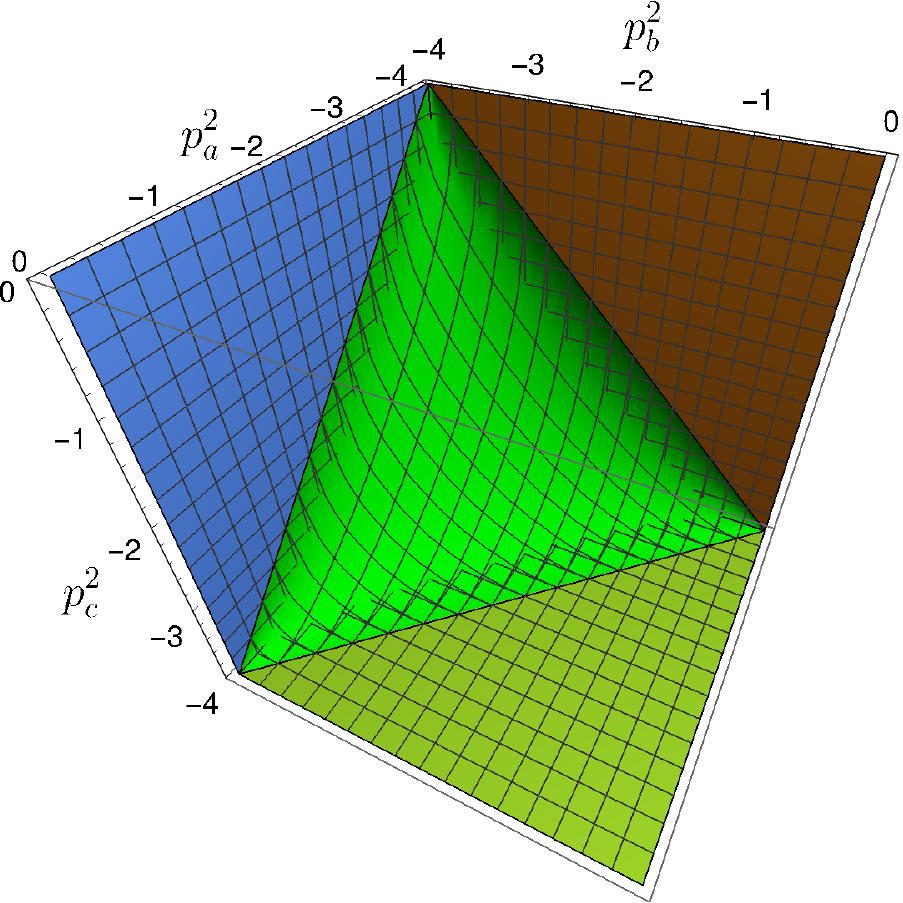}\end{center}
\caption{Full solution for thresholds of the triangle diagram including contracted diagrams.}
\label{fig:threshold_3p}
\end{figure}

It remains to determine which of the two possibilities leads to the critical point closer to the origin and thus to the relevant threshold.
For the case with two propagators, one can determine the touching point to be at $\sqrt{(p_a^2+p_b^2)/2+m^2}$.
The case with three propagators, on the other hand, has the critical points at $\pm i\,m$.
Thus, they create the highest threshold if $p_a^2+p_b^2<-4m^2$, and the two propagators create them otherwise.
The final threshold surface is thus parameterized by the walls at $-4m^2$ and the surface created by $p_{c+}^2$:
\begin{align}
\left\lbrace \begin{array}{l}
p_c^2=\frac{2m^2(p_a^2+p_b^2) + p_a^2 p_b^2 + \sqrt{p_a^2(4m^2+p_a^2)} \sqrt{p_b^2(4m^2+p_b^2)}}{2m^2} \\  \qquad \text{for} \quad -4m^2\leq p_a^2,p_b^2 \leq 0, \quad \text{and} \quad p_a^2+p_b^2\leq-4m^2  \\ ~\\
p_a^2=p_b^2=p_c^2=-4m^2 \qquad \text{else} .
\label{eq:Vert_LC_full}
\end{array} \right. 
\end{align}
The resulting surfaces are shown in \fref{fig:threshold_3p}.

We close this section with the remark that  the case of three different masses can also be analyzed in the same way \cite{Huber:2022nzs}.

\section{Conclusions}

Contour deformations are a powerful tool to access the time-like region of correlation functions with functional methods.
It encompasses the perturbative case for which the original results of the Landau analysis are recovered but can also be applied nonperturbatively.
We have applied this method to three-point functions to extract their thresholds.
In particular, we also find distinct cases depending on how many propagators are involved in the creation of the branch point reflecting the case of contracted diagrams of the Landau analysis.
The method was applied in Ref.~\cite{Huber:2022nzs} to the system of propagator and vertex Dyson-Schwinger equations of $\phi^3$ theory.
For the propagator, we could extract both the pole mass shifted by the interactions as well as the branch point.
The latter fulfills the Landau condition when the perturbative mass is replaced by the dynamical one.
For the vertex, we also confirmed the validity of the Landau conditions in a nonperturbative calculation.
Future applications are the three-point functions of quantum chromodynamics which are well-studied with functional equations but only for space-like momenta \cite{Huber:2012zj,Aguilar:2013xqa,Blum:2014gna,Eichmann:2014xya,Williams:2014iea,Williams:2015cvx,Aguilar:2016lbe,Cyrol:2017ewj,Huber:2018ned,Aguilar:2018csq,Aguilar:2019jsj,Huber:2020keu,Gao:2021wun,Pawlowski:2022oyq,Papavassiliou:2022wrb}.

\vspace{6pt} 

\authorcontributions{Conceptualization, R.A. and M.Q.H.; software, M.Q.H. and W.J.K.; investigation, R.A., M.Q.H. and W.J.K.; writing---original draft preparation, M.Q.H.; writing---review and editing, R.A. and W.J.K.; visualization, M.Q.H. All authors have read and agreed to the published version of the manuscript.}

\funding{This work was supported by the DFG (German Research Foundation) grant FI 970/11-2 and by the BMBF under contract No. 05P21RGFP3.}



\conflictsofinterest{The authors declare no conflict of interest.} 


\begin{adjustwidth}{-\extralength}{0cm}

\reftitle{References}



\bibliography{literature_anStruct3pt_ACHT2021}

\end{adjustwidth}
\end{document}